# Challenges and Characteristics of Intelligent Autonomy for Internet of Battle Things in Highly Adversarial Environments

**Alexander Kott**

U.S. Army Research Laboratory, Adelphi, MD, USA
alexander.kott1.civ@mail.mil

**Abstract**

Numerous, artificially intelligent, networked things will populate the battlefield of the future, operating in close collaboration with human warfighters, and fighting as teams in highly adversarial environments. This paper explores the characteristics, capabilities and intelligence required of such a network of intelligent things and humans – Internet of Battle Things (IOBT). It will experience unique challenges that are not yet well addressed by the current generation of AI and machine learning.

## Introduction

Internet of Intelligent Battle Things is the emerging reality of warfare. A variety of networked intelligent systems – things – will continue to proliferate on the battlefield, where they will operate with varying degrees of autonomy. Intelligent things will not be a rarity but a ubiquitous presence on the future battlefield. (Scharre 2014)

Most of such intelligent things will not be too dissimilar from the systems we see on today's battlefield, such as unattended ground sensors, guided missiles (especially the fire-and-forget variety) and of course the unmanned aerial systems (UAVs). They will likely include physical robots ranging from very small size (such as an insect-scale mobile sensors) to large vehicle that can carry troops and supplies. Some will fly, others will crawl or walk or ride. Their functions will be diverse. Sensing (seeing, listening, etc.) the battlefield will be one common function. Numerous small, autonomous sensors can cover the battlefield and provide an overall awareness to the warfighters that is reasonably compete and persistent (Fig. 1).

Other things might acts as defensive devices, e.g., autonomous active protection systems (Freedberg 2016). Finally, there will be munitions that are intended to impose physical or cyber effects on the enemy. These will not be autonomous. Instead, they will be controlled by human warfighters. This assumes that the combatants of that future battlefield will comply with a ban on offensive autonomous weapons beyond meaningful human control. Although the US Department of Defense already imposes strong restrictions on autonomous and semi-autonomous weapon systems (Hall 2017), nobody can predict what other countries might decide on this matter.

In addition to physical intelligent things, the battlefield – or at least the cyber domain of the battlefield -- will be populated with disembodied, cyber robots. These will reside within various computers and networks, and will move and acts in the cyberspace. Just like physical robots, the cyber robots will be employed in a wide range of roles. Some will protect communications and information (Stytz et al. 2005) or will fact-check, filter and fuse information for cyber situational awareness (Kott et al. 2014). Others will defend electronic devices from effects of electronic warfare. These defensive actions might include creation of informational or electromagnetic deceptions or camouflage. Yet others will act as situation analysts and decision advisers to the humans or physical robots. In addition to these defensive or advisory roles, cyber robots might also take on more assertive functions, such as executing cyber actions against the enemy systems (Fig. 2).

In order to be effective in performing these functions, battle things will have to collaborate with each other, and also with the human warfighters. This will require a significant degree of autonomous self-organization; and also of accepting a variety of relations between things and humans, e.g., from complete autonomy of an unattended ground sensor to a tight control of certain systems, and these modes will have to change flexibly as needed. Priorities, objectives, and rules of engagement will change rapidly, and intelligent things will have to adjust accordingly (Kott et al. 2016).

Clearly, these requirements imply a high degree of intelligence on part of the things. Particularly important is the necessity to operate in a highly adversarial environment, i.e., intentionally hostile and not merely randomly dangerous world. The intelligent things will have to constantly think about an intelligent adversary that strategizes to deceive and defeat them. Without this adversarial intelligence, the battle may appear inexplicable to the humans. Direct control of such intelligent things becomes impossible or limited to cases of decisions whether to take a specific destructive action.

On the other hand, humans complicate the life for intelligent things. Human and things think differently. Intelligent things, in the foreseeable future, will be challenged in un-things will not survive long enough to be useful.

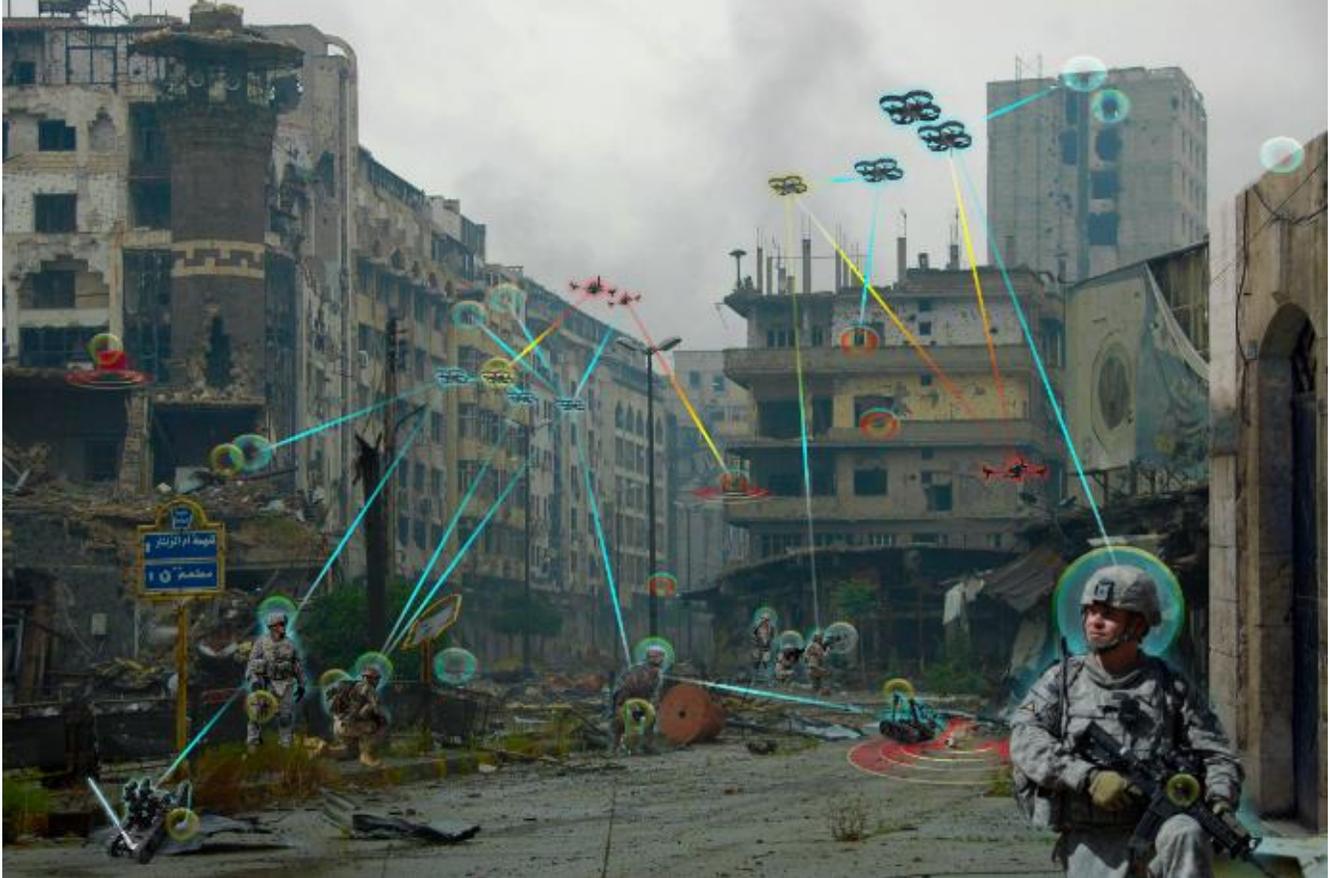

*Figure 1 Networked teams of intelligent things and humans will operate in extremely complex, challenging environment: unstructured, unstable, rapidly changing, chaotic, rubble-filled, adversarial and deceptive.*

## The Challenges of Autonomous Intelligence on the Battlefield

The vision – or rather the emerging reality -- of the battlefield populated by intelligent things, portends a multitude of profound challenges. While use of AI for battlefield tasks has been explored on multiple occasions, e.g., (Rasch et al. 2002), and AI makes things individually and collectively more intelligent, it also makes the battlefield harder to understand and to manage. Human warfighters have to face a much more complex, more unpredictable world where things have the mind of their own and perform actions that derstanding and anticipating human intent, goals, lines of reasoning and decisions. Humans and things will remain largely opaque to each other. And yet, things will be expected to perceive, reason and act while taking into account the social, cognitive and physical needs of their human teammates. Furthermore, things will often deal with humans who are experiencing extreme physical and cognitive stress, and therefore may behave differently from what can be assumed from observing humans under more benign conditions

An intelligent thing will need to deal with a world of astonishing complexity. The sheer number and diversity of things – and humans – within the IoBT will be enormous. The number of connected things, for example within a future Army brigade, is likely to be several orders of magnitude greater than in current practice. This, however, is just the beginning. Consider that intelligent things belonging to such a brigade will inevitably interact – willingly or unwillingly -- with things owned and operated by other parties, such as those of the adversary or owned by the surrounding civilian population. If the brigade operates in a large city, where each apartment building can contains thousands of things, the overall universe of connected things grows to enormous numbers. Million things per square kilometer is not an unreasonable expectation (Fig. 2).

The above scenario also points to a great diversity of things within the overall environment of the battlefield. Things will come from different manufacturers, with different designs, capabilities, and purposes, configured or machine-learned differently, etc. No individual thing will be able to use pre-conceived (pre-programmed, pre-learned, etc.) assumptions about behaviors or performance of other things it meets on the battlefield. Instead, behaviors and characteristics will have to be learned and updated autonomously dynamically during the operations. That includes humans – yes, humans are a specie of things, in a way – and therefore the behaviors and intents of humans, such as friendly warfighters, adversaries, and civilians and so on, will have to be continually learned and inferred.

The cognitive processes of both things and humans will be severely challenged in this environment of voluminous and heterogeneous information. Rather than the communications bandwidth, the cognitive bandwidth may become the most severe constraint. Both humans and things seek information that is well-formed, reasonably sized, essential in nature, and highly relevant to their current situation and mission. Unless information is useful, it does more harm than good. The trustworthiness of the information and the value of information arriving from different sources, especially other things, will be highly variable and generally uncertain. For any given intelligent thing, the incoming information could contain mistakes, erroneous observations or conclusions made by other things, or intentional distortions – deceptive information – produced by an adversary malware residing on friendly things or otherwise inserted into the environment. Both humans and things are susceptible to deception, and humans are likely to experience cognitive challenges when surrounded by opaque things that might be feeding the humans with untrustworthy information (Kott and Alberts 2017).

This reminds us that the adversarial nature of the battlefield environment is a concern of exceptional importance, above all others. The intelligent things will have to deal with an intelligent, capable adversary. The adversary will apply to things physical destruction, either by means such as gunfire, also known as "kinetic" effects, or by using directed energy weapons. The adversary will be jamming the channels of communications between things, and between things and humans. The adversary will deceive things by presenting them with misleading information. Recent research in adversarial learning comes to mind in this connection (Papernot et al. 2016). Perhaps most dangerously, the adversary will attack intelligent things by depositing malware on them.

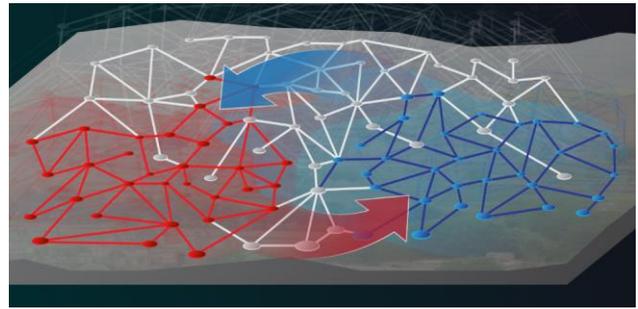

*Figure 2 Networks of the opponents will fight each other with cyber and electromagnetic attacks of great diversity and volume; most of such offensive and defensive actions will be performed by autonomous cyber agents.*

## AI will Fight the Cyber Adversary

A key assumption that must be taken regarding the IoBT is that in a conflict with a technically sophisticated adversary, IoBT will be a heavily contested battlefield (Kott 2015). Enemy software cyber agents -- malware -- will infiltrate our network and attack our intelligent things. To fight them, things will need artificial cyber hunters - intelligent, autonomous, mobile agents specialized in active cyber defense and residing on IoBT.

Such agents will stealthily patrol the networks, detect the enemy malware while remaining concealed, and then destroy or degrade the enemy malware. They will do so mostly autonomously, because human cyber experts will be always scarce on the battlefield. They will be adaptive because the enemy malware is constantly evolving. They will be stealthy because the enemy malware will try to find and kill them. At this time, such capabilities do not exist but are a topic of research (Theron et al. 2018)). Here, let's explore the desired characteristics of an intelligent autonomous agent operating in the context of IoBT.

We consider a thing – a simple senor or a complex military vehicle -- with one or more computers residing on the thing. Each computer contributes considerably into operation of the thing or systems installed on the thing. One or more of the computers are assumed to have been compromised, where the compromise is either established as a fact, or is suspected.

Due to the contested nature of the communications environment (e.g., the enemy is jamming the communications or radio silence is required to avoid detection by the enemy), communications between the thing and other elements of the friendly force can be limited and intermittent. Under some conditions, communications are entirely impossible.

Given the constraints on communications, conventional centralized cyber defense is infeasible. (Here centralized cyber defense refers to an architecture where local sensors send cyber-relevant information to a central location, where highly capable cyber defense systems and human analysts detect the presence of malware and initiate corrective actions remotely). It is also unrealistic to expect that the human war-fighters in the vicinity of the thing (if such exist) have the necessary skills or time available to perform cyber defense functions for that thing.

Therefore, cyber defense of the thing and its computing devices has to be performed by an intelligent, autonomous software agent. The agent (or multiple agents per thing) would stealthily patrol the networks, detect the enemy agents while remaining concealed, and then destroy or degrade the enemy malware. The agent has to do so mostly autonomously, without support or guidance of a human expert.

In order to fight the enemy malware deployed on the friendly thing, the agent often has to take destructive actions, such as deleting or quarantining certain software. Such destructive actions are carefully controlled by the appropriate rules of engagement, and are allowed only on the computer where the agent resides. The agent may also be the primary mechanism responsible for defensive cyber maneuvering (of which mobbing target defense is an example), deception, e.g., redirection of malware to honeypots (De Gaspari et al. 2016), self-healing, e.g., (Azim et al. 2014), and other such autonomous or semi-autonomous behaviors (Jajodia et al. 2011).

The actions of the agent, in general, cannot be guaranteed to preserve availability of integrity of the functions and data of friendly computers. There is a risk that an action of the agent may "break" the friendly computer, disable important friendly software, or corrupt or delete important data. This risk, in a military environment, has to be balanced against the death or destruction caused by the enemy if the agent's action is not taken.

Provisions are made to enable a remote or local human controller to fully observe, direct and modify the actions of the agent. However, it is recognized that human control is often impossible, especially because of intermittent communications. The agent, therefore, is able to plan, analyze and perform most or all of its actions autonomously. Similarly, provisions are made for the agent to collaborate with other agents (who reside on other computers); however, in most cases, because the communications are impaired or observed by the enemy, the agent operates alone.

The enemy malware, and its capabilities and techniques, evolve rapidly. So does the environment in general, together with the mission and constraints that the thing is subject to. Therefore, the agent is capable of autonomous learning.

Because the enemy malware knows that the agent exists and is likely to be present on the computer, the enemy malware seeks to find and destroy the agent. Therefore, the agent possesses techniques and mechanisms for maintaining a degree of stealth, camouflage and concealment. More generally, the agent takes measures that reduce the probability that the enemy malware may detect the agent. The agent is mindful of the need to exercise self-preservation and self-defense.

It is assumed here that the agent resides on a computer where it was originally installed by a human controller or by an authorized process. We do envision a possibility that an agent may move itself (or move a replica of itself) to another computer. However, such propagation is assumed to occur only under exceptional and well-specified conditions, and to take place only within friendly network – from one friendly computer to another friendly computer. This brings to mind the controversy about "good viruses." Such viruses have been proposed, criticized and dismissed earlier (Muttik 2016). These criticisms do not apply here. This agent is not a virus because it does not propagate except under explicit conditions within authorized and cooperative nodes. It is also used only in military environments where most usual concerns do not apply.

## AI will have to Advance Significantly

Agents will have to become useful team-mates – not tools -- of human warfighters on a highly complex and dynamic battlefield. Consider Fig. 1 that depicts an environment in which a highly-dispersed team of human and intelligent agents (including but not limited to physical robots) is attempting to access a multitude of highly heterogeneous and uncertain information sources, and use them for forming situational awareness and making decision (Kott et al. 2011), all while trying to survive extreme physical and cyber threats. They must be effective, in this unstructured, unstable, rapidly changing, chaotic, rubble-filled adversarial environments; learning in real-time, under extreme time constraints, with only a few observations that are potentially erroneous, of uncertain accuracy and meaning, or even intentionally misleading and deceptive (Fig. 3).

Clearly, it is far beyond the current state of AI to operate intelligently in such an environments and with such demands. In particular, Machine Learning – an area that has seen a dramatic progress in the last decade – must experience major advances in order to become relevant to the real battlefield. Let's review some of the required advances.

Learning with very small number of samples is clearly a necessity in an environment where the enemy and friends change the tactics continuously, and the environment itself is highly fluid, rich with details, dynamic and changing rapidly. Furthermore, very few if any of the available samples will be labelled, or at least not in a very helpful manner.

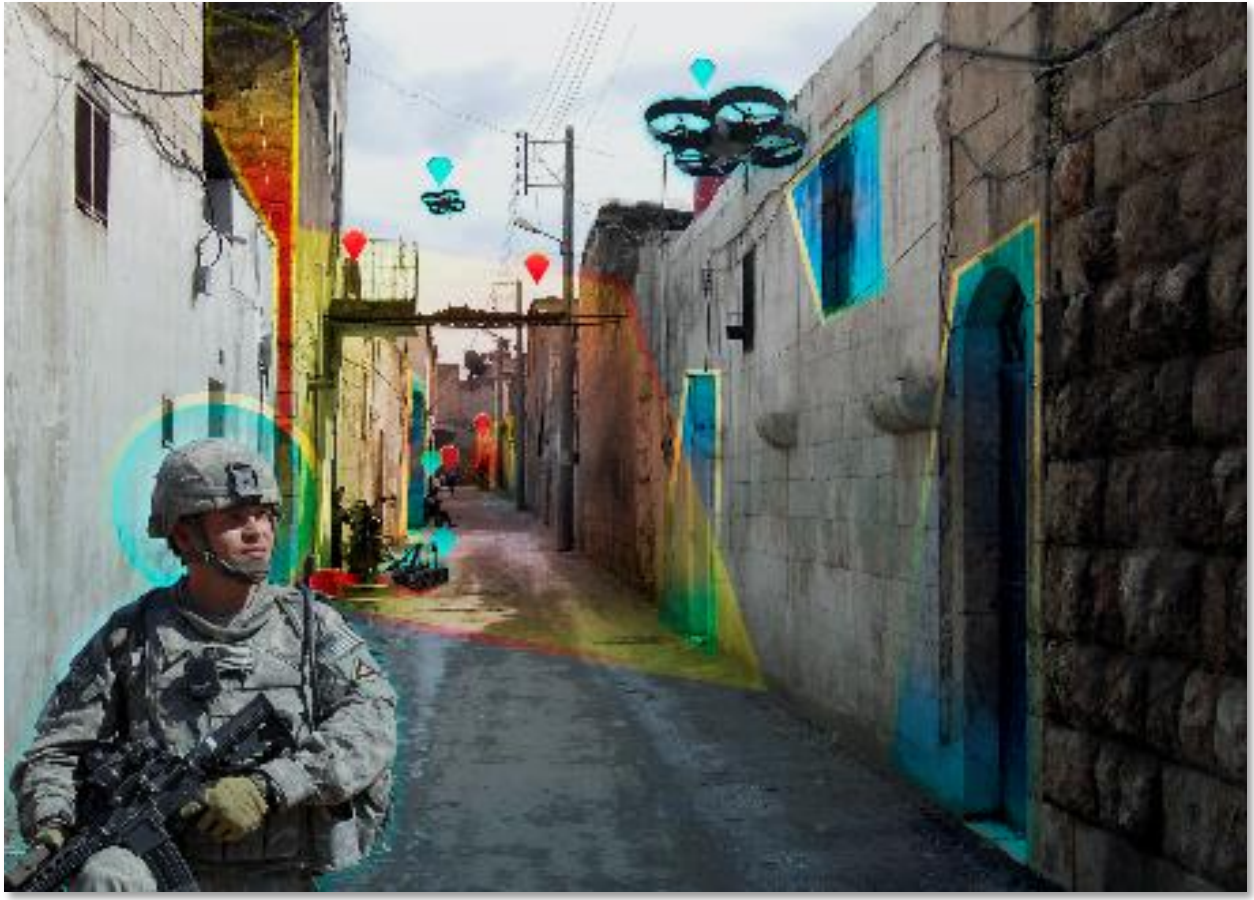

*Figure 3 . AI-enabled agents -- members of a human-agent team – will rapidly learn in ever-changing, complex environments, providing the team's commander with real-time estimates of enemy, reasoning on possible courses of action, and tactically sensible decision.*

A typical sample might be a video snippet of events and physical surroundings or a robot, for example, where the overwhelming majority of elements (e.g., pieces of rubble) are hardly relevant and potentially misleading for the purposes of learning. The information of the samples is likely to be highly heterogeneous of nature. Depending on circumstances, samples might consist of one or more of the following: still images in various parts of the spectrum (IR, visible, etc.); video; audio; telemetry data; solid models of the environment; records of communications between agents; and so on.

Some samples may be misleading in general, even if unintentionally (e.g., an action succeeds even though an unsuitable action is applied) and the machine learning algorithms will have to make the distinction between relevant and irrelevant, instructive and misleading. In addition, some of the samples might be a product of intentional deception by the enemy. In general, issues of Adversarial Learning and Adversarial Reasoning are of great importance (Papernot et al. 2016).

Yet another challenge that is uniquely exacerbated by battlefield conditions are constraints on the available electric power. Most successful AI relies on vast computing and electrical power resources including cloud-computing reach-back when necessary. The battlefield AI, on the other hand, must operate within the constraints of edge devices, such as small sensors, micro-robots, and handheld radios of warfighters. This means that computer processors must be relatively lights and small, and as frugal as possible in the use of electrical power. One might suggest that a way to overcome such limitations on computing resources available directly on the battlefield is to offload the computations via wireless communications to a powerful computing resource located outside of the battlefield.  Unfortunately, it is not a viable solution, because the enemy's inevitable interference with friendly networks will limit the opportunities for use of reach-back computational resources.

A team that includes multiple warfighters and multiple artificial agents must be capable of distributed learning and reasoning. Besides distributed learning, these include such challenges as: multiple decentralized mission-level task allocation; self-organization, adaptation, and collaboration;

space management operations; and joint sensing and perception. Commercial efforts to date have been largely limited to single platforms in benign settings. Military-focused programs like the MAST CTA (Piekarski et al. 2017), have been developing collaborative behaviors for UAVs. Ground vehicle collaboration is challenging and is largely still at the basic research level at present. In particular, to address such challenges, a new collaborative research alliance called Distributed and Collaborative Intelligent Systems and Technology (DCIST) has been initiated (https://dcist-cra.org/). Note that the battlefield environment imposes yet another complication: because the enemy interferes with communications, all this collaborative, distributed AI must work well even with limited, intermittent connectivity.

## Humans in the Ocean of Things

In this vision of the future warfare, a key challenge is to enable autonomous systems and intelligent agents to effectively and naturally interact across a broad range of warfighting functions. Human-agent collaboration is an active ongoing research area. It must address such issues as trust and transparency, common understanding of shared perceptions, and human-agent dialog and collaboration.

One seemingly relevant technology is Question Answering—the system's ability to respond with a relevant, correct information to a clearly stated question. Successes of commercial technologies of question-answering are indisputable. They work well for very large, stable, and fairly accurate volumes of data, e.g., encyclopedias. But such tools don't work for rapidly changing battlefield data, also distorted by adversary's concealment and deception. They cannot support continuous, meaningful dialog in which both warfighters and artificially intelligent agents develop shared situational awareness and intent understanding. Research is being performed to develop human-robotic dialog technology for warfighting tasks, using natural voice, which is critical for reliable battlefield teaming.

A possible approach to developing the necessary capabilities – both human and AI – is to train a human-agent team in immersive artificial environments. This requires building realistic, intelligent entities in immersive simulations. Training (for humans) and learning (for agents) experiences must exhibit high degree of realism to match operational demands. Immersive simulations for human training and machine learning must have physical and sociocultural interactions with high fidelity and realistic complexity of the operational environment. These include realistic behaviors of human actors (friendly warfighters, enemies, non-combatants), and interactions and teaming with robots and other intelligent agents. In today's video games, these interactions are limited and not suitable for simulating real battlefield. Advances in AI are needed to drive the character behaviors that are truly realistic, diverse, and intelligent.

To this end, some of the cutting-edge efforts in computer-generation of realistic virtual characters are moving towards what would be needed to enable realistic interactions in an artificial immersive battlefield. For example, Hollywood studios on a number of occasions sought technologies of the Army-sponsored Institute for Creative Technologies (http://ict.usc.edu/)to create realistic avatars of actors. These technologies enable film creators to digitally insert an actor into scenes, even if that actor is unavailable, much older or younger, or deceased. That's how actor Paul Walker was able to appear in "Fast and Furious 7," even though he died partway into filming (CBS News 2017).

## Summary

Intelligent things – networked and teamed with human warfighters – will be a ubiquitous presence on the future battlefield. Their appearances, roles and functions will be highly diverse. The artificial intelligence required for such things will have to be significantly greater than what is provided by today's AI and machine learning technologies. Adversarial – strategically and not randomly dangerous -- nature of the battlefield is a key driver of these requirements. Complexity of the battlefield – including the complexity of collaboration with humans – is another major driver. Cyber warfare will assume a far greater importance, and it will be AI that will have to fight cyber adversaries. Major advances in areas such as adversarial learning and adversarial reasoning will be required. Simulated immersive environments may help to train the humans and to train AI.

## Disclaimers